# Automatic Generation of Vectorized Montgomery Algorithm

Lingchuan Meng

Drexel University, Philadelphia PA 19104, USA
lm433@cs.drexel.edu

**Abstract.** Modular arithmetic is widely used in crytography and symbolic computation. This paper presents a vectorized Montgomery algorithm for modular multiplication, the key to fast modular arithmetic, that fully utilizes the SIMD instructions. We further show how the vectorized algorithm can be automatically generated by the SPIRAL system, as part of the effort for automatic generation of a modular polynomial multiplication library.

## 1 Introduction

The performance of the underlying modular arithmetic directly affects the performance of many crytography and computer algebra applications[8, 2, 3, 7]. Despite the rapid advancement in microprocessors, modular arithmetic still requires dedicated implementations that adapt to various hardware features, based on the comparisions of fast modular algorithm candidates.

Due to the expensive integer division in the naive implementation of modular multiplication, the main operation of modular arithmetic, fast algorithms have been developed to improve the performance by replacing the division with a sequence of less expensive operations. Barrett algorithm[1] precomputes an approximation of the inverse of the modulo, which enables a close approximation to the quotient of the product divided by the modulo. Montgomery algorithm[17] is similar to Barrett algorithm, in that it also requires precomputation with $P$. But the algorithm gains further efficiency by changing the representation to the so-called residue class. The algorithm also replaces the expensive integer division by $P$ with bitwise operations such as shift and masking.

Modular arithmetic has been implemented with some optimizations recently. For polynomial arithmetic, [7] provides efficient hand-optimized implementation of the Montgomery algorithm, and further exploits optimizations when the modulo is a Fourier prime[6]. [4] compares the algorithm candidates for modular arithmetic on graphics processing units (GPU). [20] also contains hand-optimized implementations of modular arithmetic in its standard libraries for large numbers, polynomials, etc. based on FFT and other fast algorithms.

A preliminary version of the vectorized algorithm and automatic generation mechanism were first developed and used in [16, 15] for generating fixed-size implementations of small and medium modular FFT. [11] further extends the algorithm in the automatic generation of a general-size library for a wider range of problem sizes. The more involved truncated Fourier transform (TFT)[12], the inverse TFT [14], and the modular polynomial multiplication [13, 10] have also adopted the vectorized algorithm and technique in the automatic generation of the general-size parallel libraries, respectively. The paper is based on the content of Chapter 3 in [9].

The rest of the paper is oragnized as follows. In Section 2, we introduce the background, including the analysis and examples of hardware support for integer arithmetic. Section 3 first surveys the existing algorithms for modular multiplication, including some lesser known optimization techniques, then introduces the vectorized Montgomery algorithm. The extensions to the SPIRAL sytem[19] to enable the automatic generation are explained in Section 4. The conclusion with discussion on the performance improvement is presented in Section 5.

## 2 Background

Modular arithmetic occurs in many algorithms for symbolic computation. However, the cost-effective design philosophy derived by the mainstream CPU manufacturers tends to find the *integer divide* hardware costly from both physical size and performance perspectives, which makes adding multiple units to a core prohibitive. This compromise leads to the long latency and low throughput of integer divisions. Considerable amount of effort has been made by the leading CPU manufacturers such as INTEL® and AMD® to improve the integer division algorithms in their hardware designs during the recent decades, but the modular arithmetic operations are still comparably slower than their floating-point counterparts. As a result, dedicated implementations of the modular arithmetic are required, since the performance gains from low-level operations will produce improvements in a wide variety of problems.

A data level parallelism exploited by most of the modern CPUs is the SIMD (single instruction, multiple data) parallelism. The SIMD instructions operate on vectors of data, therefore converting scalar instructions to the equivalent SIMD instructions is referred as vectorization in this paper. Figure 1 shows a typical SIMD operation, in this case a $v$-way vector multiplication, where the single instruction **VMUL** is applied to multiple data from the input vectors A and B.

The first widely-deployed desktop SIMD was Intel's MMX extension to the x86 architecture, unofficially known as the "multimedia extension". Since then, the majority of the development effort by the CPU manufacturers has been devoted to multimedia processing, hence to the floating-point SIMD extensions. As a result, integer SIMD extensions are often delayed, altered or even missed when the manufactures face the design constraints.

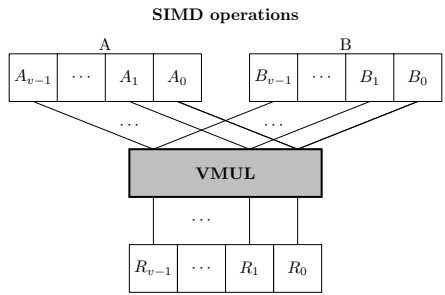

**Fig. 1.** A typical SIMD operation on $v$-way vector operands

As a warm-up, we show the complexity of multiplying two unsigned 32-bit integers and then gathering the high 32-bit and low 32-bit of the products using the existing SIMD operations. This operation is an important step in the vectorized Montgomery algorithm [17]. Unlike the vector multiply of floating-point and double precision numbers, the integer vector multiply produces a product that needs twice the width of the multiplier ($32 \rightarrow 64$).

Figure 2, 3, and 4 show three strategies for the same operation, where the light-gray shaded area denotes the low 32-bit and gray shaded area denotes the high 32-bit. All three strategies use _mm_mul_epu32 to produce the 64-bit products $T_i, i \in \{0, 1, 2, 3\}$, where $T_2$ and $T_0$ are obtained by multiplying vectors $A$ and $B$, and $T_3$ and $T_1$ are obtained by multiplying the vectors $A$ and $B$ right-shifted by 32-bit.

Figure 2 uses the floating-point shuffle instruction _mm_shuffle_ps which, controlled by an 8-bit immediate data, shuffles *two floating point vectors* into one with a latency of 1 and a throughput of 1. Type casting is required between the single-precision floating point vector type (__m128) and integer vector type (__m128i) before and after the shuffle. However, the type casting is handled by the compilers, and does not incur any overhead. Therefore, with 2 shuffles, we can gather all the high 32-bits and low 32-bits into two vectors, respectively, albeit the elements in the two vectors are out of order. Then, we can apply the integer shuffle instruction _mm_shuffle_epi32 on each of the two vectors to recover the correct order. Note that _mm_shuffle_epi32 can only shuffle integer elements *within one vector*, with a latency of 1 and throughput of 0.5 (except for on Haswell architecture where the throughput is 1). The disparity between the seemingly similar _mm_shuffle_ps and _mm_shuffle_epi32 also indicates that the floating-point vectorization strategies cannot always be naively migrated to integer vectorization.

Figure 3 uses only the integer vector instructions. First, the vectors of 64-bit products are each shuffled by _mm_shuffle_epi32 to gather the high 32-bits and low 32-bits within each vector. Then, _mm_unpacklo_epi32 and _mm_unpackhi_epi32 interleave the 32-bits within the low 64-bits and high 64-bits, respectively, of the two shuffled vectors to gather the all the low 32-bits and high 32-bits in the cor-rect order. Note that `_mm_unpacklo_epi32` and `_mm_unpackhi_epi32` both have a latency of 1 and throughput of 0.5 (except for on Haswell architecture where the throughput is 1).

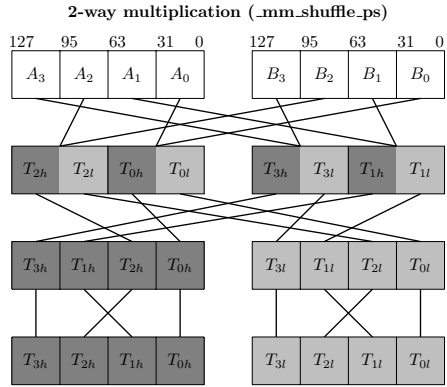

**Fig. 2.** 2-way vectorized multiplication with floating point shuffle and cast

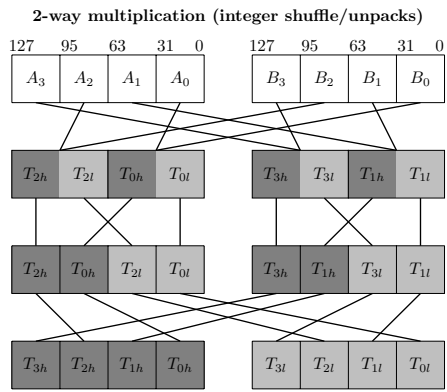

**Fig. 3.** 2-way vectorized multiplication with integer shuffle and unpack

Figure 4 uses the integer blend instruction available in AVX2. The vector containing 64-bit products $T_2$ and $T_0$ is shuffled with `_mm_shuffle_epi32` to flip the high and low 32-bits within each product, resulting in the correct interleaved order of the 4 high 32-bits in the two product vectors. Next, `_mm_blend_epi32` is applied to the product vectors twice with different 8-bit immediate data to gather the high 32-bits in-order and low 32-bits out-of-order. The vector containing the low 32-bits is shuffled again to recover the correct order. The `_mm_blend_epi32` is only available in the Haswell architecture with a latency of 1 and a throughput of 0.33.

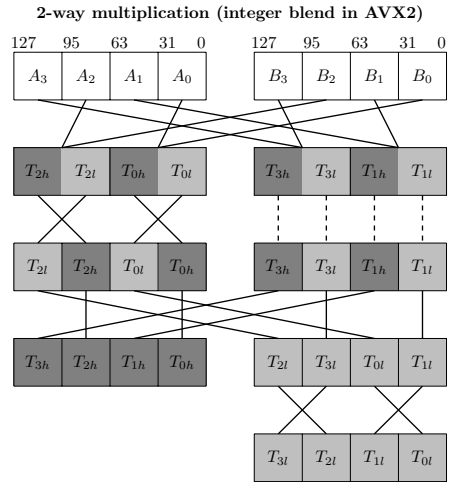

**Fig. 4.** 2-way vectorized multiplication with integer blend

Figure 2, 3, and 4 show a typical scenario of integer vectorization. The integer SIMD instructions made available under the design constraints often overlap with each other in terms of functionality, causing confusions in deciding the best implementation strategy. They also often work differently than the seemingly similar floating point counterparts, preventing the floating point vectorization strategies from being directly migrated to integer vectorization. When the functionality is identical, the integer vector instructions often tend to have higher latency or lower throughput compared to their floating point counterparts. These limitations make vectorizing the fast modular arithmetic algorithms challenging.

## 3 Fast Algorithms

In this section, we review the fast algorithms for modular arithmetic operations, with a particular focus on the pivotal modular multiplication, since the modular addition and subtraction are straightforward. The fast algorithms operate over the finite fields defined by word-size primes. A key assumption of the fast algorithms is that arithmetic with powers of two is highly efficient, which is true on modern hardware. The arithmetic operations modulo a power of two can be performed extremely quickly with bitwise shifts and logical instructions. Note that the algorithms, namely the Barrett algorithm and Montgomery algorithm, are well-known and implemented as scalar code in some existing libraries, but the vectorized implementations were first devised in [16].

### 3.1 Existing Algorithms

**Barrett algorithm**[1] precomputes an approximation of the inverse of $P$, which enables a close approximation to the quotient of $ab$ divided by $P$. Suppose $\lceil \log_2(P) \rceil = k$, i.e., $P$ can be represented with $k$ valid bits. Let $P' = \lfloor 2^{2k}/P \rfloor$, a rescaled numerical inverse of $P$ stored as a precomputation. The Barret algorithm computes $ab \bmod P$ as follows:

---
**Algorithm 1** Barrett Algorithm
---
  **function** BARRETT$(a, b)$
    $m \leftarrow \lfloor ab/2^k \rfloor$
    $q \leftarrow \lfloor mP'/2^k \rfloor$
    $t \leftarrow ab - qP$
    **while** $t \geq P$ **do**
      $t \leftarrow t - P$
    **end while**
    **return** $t$
  **end function**

---

The value of $t$ before the while loop is less than $4P$, therefore the while loop executes at most 3 times. Hence the total arithmetic operations consist of three multiplications, at most four subtractions, and two bitwise shifts that perform the integer division by a power of two.

**Montgomery algorithm**[17] is similar to Barrett algorithm, in that it also requires precomputation with $P$. But the algorithm gains further efficiency by changing the representation to the so-called residue class. The algorithm also replaces the expensive integer division by $P$ with bitwise operations such as shift and masking.

For $P > 1$, the algorithm defines an $P$-*residue* to be a residue class modulo $P$. First, the algorithm selects a $R$ coprime to $P$ such that $R > P$ and computations modulo $R$ are inexpensive to process. Let $R^{-1}$ and $P'$ be integers satisfying $0 < R^{-1} < P$ and $0 < P' < R$ and $RR^{-1} - PP' = 1$, which can be computed using the extended Euclieand algorithm. For $i \in \mathbb{Z}_P$, the residue class $\bar{i}$ is defined as $iR \bmod P$. With this residue system, we can quickly compute $\bar{a}\bar{b}R^{-1} \bmod P$ from $\bar{a}$ and $\bar{b}$ if $0 \leq \bar{a}\bar{b} < RP$, as shown in the algorithm below:

The correctness of the algorithm is shown in the comments lead by ▷. Given two numbers $x, y \in \mathbb{Z}_P$, let $z = $REDC$(\bar{x}, \bar{y})$. Then $z = \bar{x}\bar{y}R^{-1} \bmod P \equiv (xR)(yR)R^{-1} \bmod P \equiv xyR \bmod p$. Also, $0 \leq z < P$, so $z$ is the product of $x$ and $y$ in the residue representation. Note that the change of representation does not affect the modular addition and subtraction algorithms.

The total arithmetic operations consist of three integer multiplications, but fewer addition/subtractions compared to Barrett algorithm. As we later vectorize the algorithm, we further see that the second multiplication is a "short product" where only the low-order bits are needed. This observation is a key to efficient adaptation to the available integer vector instructions. The overhead

**Algorithm 2** Montgomery Algorithm

**function** REDC($\bar{a}, \bar{b}$)
    $T \leftarrow \bar{a}\bar{b}$
    $m \leftarrow (T \mod R)P' \mod R$                                 ▷ $m \equiv \bar{a}\bar{b}P' \mod R$
    $t \leftarrow (T + mP)/R$        ▷ $mP \equiv \bar{a}\bar{b}P'P \equiv -\bar{a}\bar{b} \mod R \Rightarrow t \equiv \bar{a}\bar{b}R^{-1} \mod P$
    **if** $t \geq P$ **then**                            ▷ $0 \leq \bar{a}\bar{b} + mP < RP + RP \Rightarrow 0 \leq t < 2P$
        **return** $t - P$
    **else**
        **return** $t$
    **end if**
**end function**

of converting to and from the residue representation is clearly expensive, but it becomes negligible in most applications where long sequences of modular arithmetic is performed. Therefore, we will focus on the efficient implementation and integration of Montgomery algorithm in SPIRAL.

Montgomery algorithm can be optimized further when the prime $P = c2^n + 1$, also known as a Fourier prime when $c$ is small and $n$ is sufficiently big [6]. More precisely, we require $n \geq l/2$ where $l$ is the bit length of $P$ and $R = 2^l$. Machine word-size multiplication modulo such Fourier primes can be done efficiently with the algorithm below:

**Algorithm 3** Montgomery Algorithm over Fourier Primes

**function** FOURIERREDC($\bar{a}, \bar{b}$)
    $q_1 \leftarrow \bar{a}\bar{b}/R$                                                       ▷ / is the integer division
    $r_1 \leftarrow \bar{a}\bar{b} \mod R$
    $q_2 \leftarrow c2^n r_1/R$
    $r_2 \leftarrow c2^n r_1 \mod R$
    $q_3 \leftarrow c2^n r_2/R$
    $t \leftarrow q_1 - q_2 + q_3$
    $t \leftarrow t + (t \gg 31) \,\&\, P$
    $t \leftarrow t - P$
    $t \leftarrow t + (t \gg 31) \,\&\, P$
    **return** $t$
**end function**

We first show that $t \equiv \bar{a}\bar{b}R^{-1} \mod P$. By the definitions of $q_1$ and $r_1$, we have $\bar{a}\bar{b} = q_1 R + r_1$. Then we have

$$\begin{aligned}
\bar{a}\bar{b}R^{-1} &\equiv q_1 + r_1 R^{-1} \mod P \\
&\equiv q_1 - c2^n r_1 R^{-1} \mod P \quad \triangleright \text{since } c2^n \equiv -1 \mod P \\
&\equiv q_1 - q_2 - r_2 R^{-1} \mod P \quad \triangleright \text{since } q_2 = c2^n r_1/R, r_2 = c2^n r_1 \mod R \\
&\equiv q_1 - q_2 + q_3 + r_3 R^{-1} \mod P \quad \triangleright \text{since } q_3 = c2^n r_2/R, r_3 = c2^n r_2 \mod R \\
&\equiv q_1 - q_2 + q_3 \mod P \quad \triangleright \text{since } r_3 = 0
\end{aligned}$$

To understand why $r_3 = 0$, we have $r_3 = c2^n r_2 \mod R \equiv c2^n(c2^n r_1 - q_2 R) \mod R \equiv c^2 2^{2n} r_1 - c2^n q_2 R \mod R$, where both $c2^n q_2 R$ and $c^2 2^{2n} r_1$ are multiples of $R$, as $R = 2^l$ and $l \leq 2n$, resulting in $r_3 = 0$. It can be further proved that $t \in \{-(P-1), 2(P-1)\}$.

Due to the aforementioned advantage over the Barrett algorithm, the Montgomery algorithm is used throughout the library generation for modular polynomial arithmetic in SPIRAL. Next, we introduce the vectorized Montgomery algorithm that utilizes the available SIMD instructions for improved data-level parallelism.

### 3.2 Vectorized Montgomery Algorithm

Vectorization of the Montgomery algorithm is more complicated compared to the vectorization of modular addition and subtraction. Given the size $l$ of the multipliers, e.g., a machine-word size of 32, the multiplication steps in the algorithm generate the intermediate products that require $2l$-bits to store. This observation, coupled with the availability of the integer SIMD extensions, requires the vectorized implementation to be well designed to fully utilize the available vector width for optimal performance.

Next, we show how a SIMD extension with $sl$-bit vector registers can be efficiently utilized in a mixture of $s$-way and $s/2$-way vectorization in order to vectorize the Montgomery algorithm (Algorithm 2) with $l$-bit inputs. That is, we fully exploit the vector level parallelism and only reduce to $s/2$-way when the intermediate products are needed in full. Next, we use 32-bit multipliers and SSE 4.2 with 128-bit vector registers as a running example.

Figure 5 illustrates the vectorized Montgomery algorithm which proceeds from top to bottom. The labels correspond to the variables and expressions in Algorithm 2. The solid lines denote operations on the vectors, and the dashed lines denote that the vectors are simply kept and used later. The prefix of the operators, such as 2 and 4, indicate whether the operations are 2-way or 4-way.

First, two 2-way integer multiplications are performed to produce four 64-bit $T$'s. Note that in Algorithm 2, $T$ is then reduced to $\mathbb{Z}_R$ and multiplied with $P'$, after which the product is again reduced to $\mathbb{Z}_R$ to produce $m$. Here, the first reduction is applied to the lower 32-bits of $T$'s after they are properly gathered, and the multiplication only needs to generate the low 32-bit of the product (hence the 4-way `_mm_mullo_epi32`), as the second reduction to $\mathbb{Z}_R$ only operate on the low 32-bit per the definition of $R$. The reduction to $\mathbb{Z}_R$ can be efficiently performed as a `bitwise and` with $R-1$ to produce $m$.

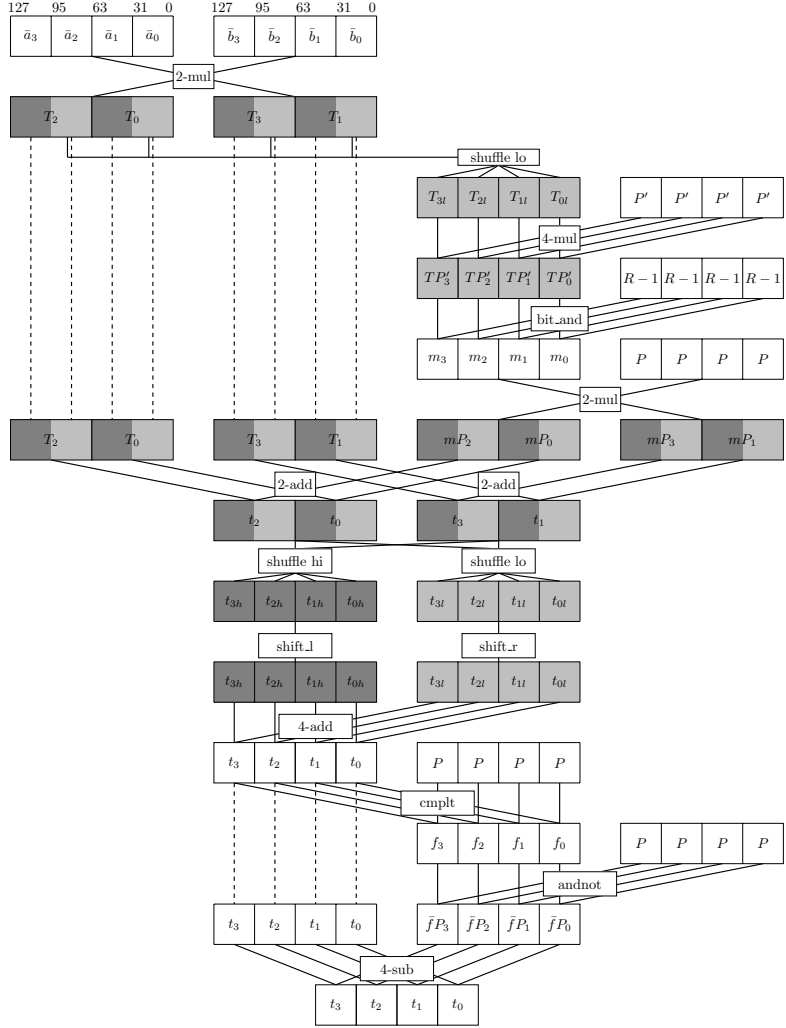

**Fig. 5.** The vectorized Montgomery algorithm

Next, $m$ is 2-way multiplied with $P$, whose product is then 2-way added to $T$. The 2-way addition is required as carries may happen across the low and high 32-bits. $t_h$'s and $t_l$'s are then gathered in-order before being divided by $R$. The division is performed separately on the high and low 32-bits. The high 32-bits is a multiple of $R$, therefore division by $R$ is equivalent to a `bitwise left shift`; The low 32-bit divided by $R$ is equivalent to a `bitwise right shift`. Both shifts are safe based on the definition of $R$. Then the shifted results are 4-way added to produce $t$'s. Finally, $t$'s are reduced to $\mathbb{Z}_P$ by selectively subtracting $P$ where $t_i \geq P$.

## 4  SPIRAL Extensions for Modular Arithmetic

SPIRAL[19] used to concentrate on the numeric code generation and optimization, as the input to the DSP algorithms and numeric kernels are either complex or real. In our problem domain of modular polynomial arithmetic, the performance of the underlying modular arithmetic is critical. In this section, we introduce the extensions to SPIRAL's limited integer arithmetic generation to support efficient modular arithmetic generation, including new data types, vector ISAs and library templates.

Data types play a central role at many levels of code generation and rewriting in SPIRAL. The built-in data types can be divided into two categories: 1. the primitive data types, such as *TReal* and *TCplx*, and 2. the composite data types, such as *TVect*. The primitive data types are associated with the operators in scalar arithmetic operations, which can be consumed by the type-based IR rewriting rules and by the corresponding unparsers. The composite data types, as the name indicates, are compositions of the primitive data types. For example, a 4-way floating point vector of is represented as *TVect*(*TReal*, 4), where 4 indicates the width of the vector.

Both the primitive and composite data types can be pairwise unified when two or more operators of different types are presented in the same arithmetic operation. Roughly speaking, the type unification rules are similar to the type conversions in languages like C when two arguments of the operation are of different types. For example:

$$a \in \mathit{TCplx}, b \in \{\mathit{TInt},\ \mathit{TUInt},\ \mathit{TReal},\ \mathit{TCplx}\}, \mathrm{Unify}(a,b) \to \mathit{TCplx},$$
$$a \in \mathit{TVect}(t_a, s_a), b \in \mathit{TVect}(t_b, s_b), \mathrm{Unify}(a,b) \to \mathit{TVect}(\mathrm{Unify}(t_a, t_b), \max(s_a, s_b)). \quad (1)$$

(1) shows that if the data type of one operator is *TCplx*, then the unified data type is *TCplx*, if the other operator is any type in {*TInt*, *TUInt*, *TReal*, *TCplx*}; If the two operators are both *TVect*, then the Unify function unifies their primitive data types and takes the longer vector length in the resulting composite type.

New modular data types and corresponding unification rules are implemented to support the generation of modular arithmetic, namely *TModInt*, *TModInt64* and *TModReal*. The integer modular types are directly associated with the 32-bit and 64-bit integers in `stdint.h`, respectively, as an effort to precisely control the integer size in code generation and optimization. On the other hand, *TModReal* is used in the algorithms where conversions between integer and floating point are required. The new unification rules enforce the modular arithmetic when at least one operator is a modular type. For example:

$$a \in \mathit{TModInt}, b \in \{\mathit{TBool},\ \mathit{TUInt},\ \mathit{TInt},\ \mathit{TModInt}\}, \mathrm{Unify}(a,b) \to \mathit{TModInt}.$$

The data types also guide the generation of compilable code when the (rewritten) IR is unparsed. Therefore the unparsers must be extended to properly generate modular code. Additionally, the library generation in SPIRAL parameterizes the library template files where many components are hard coded for real or complex arithmetic. The template files have since been extended for modular arithmetic, but the details are omitted in this paper.

SPIRAL maintains internally a collection of ISAs (instruction set architecture) that capture the essential features of the potential target platforms. The target ISA is used during the library generation process in order to produce valid and efficient code for the given platform. Table 1 shows the snapshot of a newly added ISA called "SSE 4x32m" for the vectorized modular arithmetic. As we can see, the ISA encapsulates the platform-specific features, including vector width (in terms of number of elements), the primitive data type, and short-vector memory operations, etc.

| SSE 4x32m | |
|---|---|
| info | "SSE 4-way 32-bit modular integer ISA" |
| v | 4 |
| t | $TVect(TModInt, 4)$ |
| ctype | "int32_t" |
| includes | ()$\to$ Concat("stdint.h", ...) |
| svload_init | (vt) $\to \ldots$ |
| svstore_init | (vt) $\to \ldots$ |
| ... | |

Table 1. A snapshot of a sample ISA

The scalar and vectorized Montgomery algorithms shown in Section 3 can be automatically generated by SPIRAL with the new extensions. The algorithms are encoded as rewriting rules at the IR (intermediate representation) level. Then, the rewritten IR implementing the algorithm is unparsed by a built-in unparser to generate compilable code that is valid and efficient for the target platform.

Figure 6 shows the workflow of this process using the vectorized algorithm as an example. An IR expression `assign(res, mul(a, b))` where `res` is of type $TVect(TModInt, 4)$ can be captured by the pattern defined in the type-based IR rewriting rule. The rule encapsulates the vectorized algorithm and produces the algorithm expressed as a chain of IR expressions including the declaration of temporary variables. Then, the rewritten IR expressions are unparsed by the vector unparser to produce efficient implementation of the vectorized Montgomery algorithm. Note that the scalar and vector unparsers in SPIRAL have also been expanded considerably to support the generation of integer and modular arithmetic, but the details are omitted in this paper.

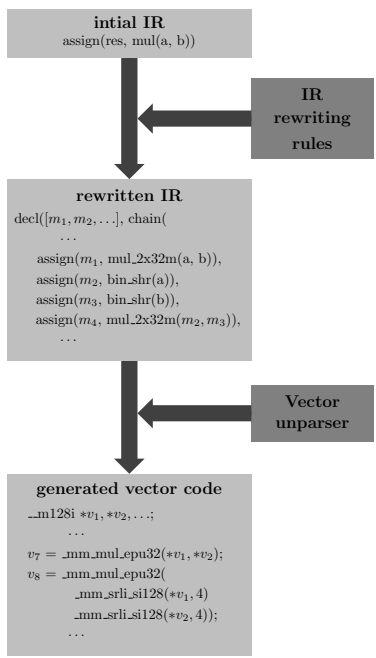

**Fig. 6.** An example of vector code generation for modular arithmetic in SPIRAL

## 5 Conclusion

The performance improvement from the automatic generation of the vectorized Montgomery algorithm and the other modular arithmetic operations can be found in the auto-tuned modular FFT libraries[16, 11]. The automatically generated libraries are an order of magnitude faster than the state-of-the-art of the hand-optimized implementations. We have also shown how SPIRAL can be extended to generate vectorized modular arithmetic. As a result, various transforms and algorithms relying on fast modular arithmetic can easily benefit from the SIMD acceleration, while being represented and optimized at a high abstraction level with little human effort.